\documentclass[a4paper,11pt]{article}

\usepackage[margin=2.5cm]{geometry}
\usepackage{url}
\usepackage[utf8]{inputenc} 
\usepackage{booktabs}
\usepackage{caption}
\usepackage{subcaption}
\usepackage{graphicx}
\usepackage{nameref}
\usepackage{colortbl}
\usepackage{xcolor}
\usepackage{multirow}

\newcommand{\NO}{white!30!cyan}
\newcommand{\SI}{white!35!red}

\providecommand{\keywords}[1]
{
  \small	
  \textbf{\textit{Keywords---}} #1
}

\begin{document}

\title{\bf Characterizing networks of propaganda on Twitter: a case study}

\author{
        Stefano Guarino$^{1}$, Noemi Trino$^{2}$,  Alessandro Celestini$^{3}$, Alessandro Chessa$^{4}$, Gianni Riotta$^{5}$ \\\\
        \small $^{1,3}$Institute for Applied Mathematics, National Research Council, Rome, Italy. \\
        \small $^{2,4,5}$Luiss “Guido Carli” University, Rome, Italy.\\
        \small $^{4}$Linkalab, Cagliari, Italy.\\\\
        Email: $^{1}$s.guarino@iac.cnr.it, $^{3}$a.celestini@iac.cnr.it
}

\date{}

\maketitle

\begin{abstract} 
The daily exposure of social media users to propaganda and disinformation campaigns has reinvigorated the need to investigate the local and global patterns of diffusion of different (mis)information content on social media.
Echo chambers and influencers are often deemed responsible of both the polarization of users in online social networks and the success of propaganda and disinformation campaigns.
This article adopts a data-driven approach to investigate the structuration of communities and propaganda networks on Twitter in order to assess the correctness of these imputations.
In particular, the work aims at characterizing networks of propaganda extracted from a Twitter dataset by combining the information gained by three different classification approaches, focused respectively on (i) using Tweets content to infer the ``polarization'' of users around a specific topic, (ii) identifying users having an active role in the diffusion of different propaganda and disinformation items, and (iii) analyzing social ties to identify topological clusters and users playing a ``central'' role in the network. 
The work identifies highly partisan community structures along political alignments; furthermore, centrality metrics proved to be very informative to detect the most active users in the network and to distinguish users playing different roles; finally, polarization and clustering structure of the retweet graphs provided useful insights about relevant properties of users exposure, interactions, and participation to different propaganda items. 
\end{abstract}


\section{Introduction}

The 2016 US presidential election veritably marked the transition from an age of `post-trust' \cite{lofstedt2005risk}, to an era of `post-truth' \cite{higgins2016post}, with contemporary advanced democracies experiencing a rise of anti-scientific thinking and reactionary obscurantism, ranging from online conspiracy theories to the much-discussed “death of expertise” \cite{nichols2017death}.
The long-standing debate about the relationship between media and public good has been reinvigorated: the initial euphoria about the ``openness'' of the Internet \cite{levy2002cyberdemocratie} has been taken over by a widespread concern that social media may instead be undermining the quality of democracy \cite{tucker2018social}.
Media outlets, public officials and activists are supplying citizens with different, often contradictory ``alternative facts'' \cite{allcott2017social}.
In this context, social media platforms would be fostering ``selective exposure to information'', with widespread diffusion of ``echo chambers'' and ``filter bubbles'' \cite{sunstein2001republic, pariser2011filter}.
Propaganda actions may be now more effective than ever, representing a major global risk, possibly able to influence public opinion enough to alter election outcomes \cite{van2017inoculating, shao2018anatomy, guess2019less}. 

As a first step towards the disruption of these networks of propaganda, researchers have been trying to model the social mechanisms that make users fall prey of partisan and low-quality information.
From a psychological point of view, news consumption is mainly governed by so-called ``informational influence'', ``social credibility'', ``confirmation bias'' and ``heuristic frequency''~\cite{shu2017fake, del2017modeling}.
This means that social media users tend to shape their attitude, belief or behavior based on arguments provided in online group discussions, using popularity as a measure of credibility, privileging information that confirms their own prior beliefs and/or that they hear regularly.
These phenomena are exacerbated by the general incapability of making good use of the great amount of available information, a problem which can be modeled relying on the dualism of information overload \emph{vs.} limited attention~\cite{qiu2017limited}, or on the principles of information theory and (adversarial) noise decoding~\cite{brody2018model}.
However, there is still a lack of evidence in the literature regarding the processes that lead to the structuration of digital ecosystems where polarized and unverified claims are especially likely to propagate virally.
Are these a natural consequence of the existence of communities with homogeneous beliefs -- \emph{i.e.}, echo chambers -- and of the organized actions of ``propaganda agents'', or are we missing a piece?

To provide a first answer to this and other related questions, the present paper takes a data-driven approach.
Specifically, we aim at demonstrating the importance of characterizing networks of propaganda on Twitter by combining the information gained by three different classification approaches: (i) using the content of tweets to determine users' ``polarization'' with respect to a main theme of interest; (ii) telling apart users having an active role in the diffusion of different propaganda and disinformation items related to that theme; (iii) analyzing social ties to identify topological clusters and users playing a ``central'' role in the network.
Our main goal is addressing the following research questions:
\begin{itemize}
    \item Is modularity-based network clustering ``stable'' or are the patterns of cohesion among users dependent of the topics of discussion? In other terms, is the exposure/participation to propaganda of a given user a direct consequence of his/her own global interactions with other users?
    \item Can we use centrality metrics for detecting users playing specific roles in the production-diffusion chain of propaganda? If yes, what metrics should we mostly rely on? And are these users ``consistently'' involved in the diffusion of related yet different propaganda items?
    \item What is the role of polarization in the analysis?  How shall we use the available information about the political/social ``goal'' of a propaganda item to enrich the graph-based analysis of the corresponding network of propaganda?
\end{itemize}
Our methodology will be applied to a case study concerning the constitutional referendum held on December 4, 2016 in Italy, by means of a dataset composed of over 1.3 millions tweets.
As a side result, we will provide insights regarding the reasons of the success of specific propaganda items and the existence of ``propaganda hubs'' and ``authorities'', \emph{i.e.}, accounts that are critical in fostering propaganda and spreading disinformation campaigns.

\subsection{Related work}
As reported by a recent Science Policy Forum article~\cite{lazer2018science}, stemming the viral diffusion of fake news largely remains an open problem.
The body of research work on fake news detection is vast and heterogeneous: linguistics-based techniques~\cite{markowitz2014linguistic, feng2012syntactic, feng2013detecting} coexist with network-based techniques~\cite{ciampaglia2015computational, papacharissi2012affective, karadzhov2017fully} as well as machine-learning-based approaches \cite{castillo2011information, zubiaga2018detection}.
Yet, (semi-)automatic debunking seems not an adequate response if considered alone~\cite{margolin2018political, shin2017partisan}.
Experimental evidence confirms the general perception that, on average, fake news get diffused farther, faster, deeper and more broadly than true news~\cite{silverman2016most}.
Users are more likely to share false and polarized information and to share it rapidly, especially when related to politics~\cite{vosoughi2018spread}, while the sharing of fact-checking content typically lags that of fake news by at least 10 hours~\cite{shao2016hoaxy}.
Furthermore, debunking is often associated to counter-propaganda and disseminated online through politically-oriented outlets, thus reinforcing selective exposure and reducing consumption of counter-attitudinal fact-checks~\cite{shin2017partisan}.
Besides the technical setbacks, the existence of the so-called ``continued influence effect of misinformation'' is widely acknowledged among socio-political scholars~\cite{skurnik2005warnings}, thus questioning the intrinsic potential of debunking in contrasting the proliferation of fake news.

In this regard, the efforts deployed by major social media platforms seem insufficient.
As of 2017, Twitter -- the most widely studied of such platforms -- expressed an alarmingly shallow stance towards disinformation, stating that bots are a ``positive and vital tool'' and that Twitter is by nature ``a powerful antidote to the spreading of false information'' where ``journalists, experts and engaged citizens can correct and challenge public discourse in seconds''~\cite{crowell2017our}. 
In the meanwhile, based on two millions retweets produced by hundreds thousands accounts in the six months preceding the 2016 US presidential election, researchers were coming to the conclusion that the core of Twitter's interaction network was nearly fact-checking-free while densely populated of social bots and fake news~\cite{shao2018spread}.

Characterizing misinformation and propaganda networks on social media thus recently emerged as a primary research trend~\cite{subrahmanian2016darpa, shao2018anatomy, bovet2019influence}. 
Data collected on social media are paramount for understanding disinformation disorders~\cite{bovet2019influence}: they are instrumental to analyze the global and local patterns of diffusion of unreliable news stories~\cite{allcott2017social} and, to a broader level, to understand the relevance of propaganda on public opinion, possibly incorporating thematic, polarity or sentiment classification~\cite{vosoughi2018spread}, thus unveiling the structure of social ties and their impact on(dis)information flows~\cite{bessi2016social}.
Investigating the relation between polarization and information spreading has also been shown to be instrumental for both uncovering the role of disinformation in a country's political life~\cite{bovet2019influence} and predicting potential targets for hoaxes and fake news~\cite{vicario2019polarization}. 
Finally, recent work used network-based features as instruments to describe, classify and compare the diffusion networks of different disinformation stories as opposed to ``main-stream'' news, making a promising step towards text-independent fake news detection~\cite{pierri2020investigating}.

A relevant issue emerging from the literature is quantifying the representativeness of data extracted from real-time social media in general, and more specifically from Twitter, when these data are used to forecast opinion trends and vote shares in elections.
In particular, the socio-demographic composition of Twitter users may be not representative of the overall population and may thus manifest different political-preferences from non-Twitter users~\cite{bakker2011good},~\cite{burckhardt2016tweet}.
This potential mismatch could be accompanied by a self-selection bias: as some scholars showed~\cite{ceron2016politics}, the largest number of comments is often produced by the more active and politically mobilitated users, while the vast majority of accounts has a limited activity~\cite{gayo2011limits}.
Nonetheless, the main goal of this paper is making one step forward in the understanding of the role of propaganda in shaping the political debate in Italy.
To this end, Twitter is extremely representative: it is in fact the reference social media in Italy to discuss political issues.
Investigating to which extent our findings may be extended to the Italian population at large is left to future work.

\section{Background}\label{sec:background}

After the crucial 2013 election, that had imposed an unprecedented tri-polar equilibrium in the Italian political scenario, the 2016 referendum determined the collapse of the entire political scene, with the defeat of the center-left ``Democratic Party'' and the successive resignation of its leader and head of government, Matteo Renzi, architect of the consultation. 
The government reform was in fact strongly defeated, with ``NO'' percentages at 59.12\% and ``YES'' at 30.88\%. 
Offline trends showed how political polarisation and divisions among party leaders fostered the grassroots activism of the YES and NO front committees, reinforcing opposite views regarding the reform. The NO faction was a composite formation supported by both left-wing and right-wing parties, with alternative yet sometimes overlapping political justifications. 
Subsequently, the 2018 elections sanctioned the major rise of two euro-skeptic and populist formations, ``5 Stars Movement'' and ``The Northern League'', who were the main actors of opposition to the 2016 referendum. 

The constitutional referendum offered to these rising parties an extraordinary window of opportunity in propaganda building, by imposing carefully selected instrumental news-frames and narratives and using social media as strategic resources for community-building and alternative agenda setting.  
Social media -- and Twitter in particular -- have in fact constituted a strategic tool for newly born political parties, that through the activation of the two-way street mediatization could incorporate their proposals into conventional media, still maintaining a critical, even conspiratorial attitude towards traditional media~\cite{alonso2018communication, schroeder2018digital}.
More generally, the dichotomous structuration of referendum offered to both political alignments the chance to align the various issues along a pro-anti/status quo spectrum. 
The cleavage was strategically used by both coalitions, which adopted opposite frames to stress their position:
\begin{itemize}
\item on the one hand, the referendum was framed as a tool of ``rottamazione'', the process of political renovation at the center of Renzi's political agenda;
\item on the other one, on the NO front, it was inserted in the broader cleavage between anti-parties and traditional parties, pointed as an expression of old interests and privileges.
\end{itemize}

\section{Data Collection}\label{sec:background}

For data collection we relied on Twitter's Streaming API, scraping tweets containing any combination of the following hashtags: ``\#ReferendumCostituzionale'', ``\#IoVotoNO'', ``\#SIcambia'', ``\#SIRiforma'', ``\#Italiachedicesì'', ``\#Italiachedicesi'', ``\#bastaunsi'', ``\#referendum'', ``\#costituzione'', ``\#riformacostituzionale'', ``\#famiglieperilno'', ``\#bastaunsì'', ``\#bastaunsi'', ``\#referendumsociali''.
These are a mix of ``trending'' hashtags, official hashtags of the referendum campaign, and popular hashtags used by the supporters of the two fronts. Data was collected for the six months preceding the referendum, that is, from July 05, 2016, to December 04, 2016, but we only consider the tweets dated from November 01 in this paper in order to focus on the most relevant part of the campaign.

\subsection{Propaganda items}
Following the literature, in order to identify the main topics and themes of disinformation of the political campaigning we relied on the activity of fact-checking and news agencies who reported lists of (dis)information news stories that went viral during the referendum campaign.
Mostly based on the work by fact-checking web portal \emph{Bufale.net}~\cite{bufale.net}, online newspaper \emph{Il Post}~\cite{ilpost}, and political fact-checking agency \emph{Pagella Politica}\footnote{Pagella Politica is partner of the EU H2020 SOMA Project.}~\cite{pagellapolitica}, we were able to identify twelve main stories, including both general theories and very specific news pieces.
To widen the scope of the analysis, we considered news, theories and topics of discussion that could be associated to information disorders in its broader sense.
This includes \emph{factual} (\emph{i.e.}, verifiably true/false) claims as well as stories (\emph{e.g.}, hearsays, rumors and conspiracy theories) that cannot be deemed true/false with certainty, with no distinction between deliberate and organized disinformation/propaganda and unintentionally propagated misinformation.

Differently from related work~\cite{pierri2020investigating} that used the presence of a specific url for collecting tweets associated to a news story of interest, we set up a custom keyword-based query in order to search our dataset for tweets that discuss a given topic in a broader sense.
For each of the twelve propaganda items considered, we thus manually selected relevant textual content related to that story -- news pieces, tweets, work of debunking agencies -- from which we extracted a suitable query. An example of such queries is the following (corresponding to what will be later denoted PI2):
\begin{quote}
    \emph{('illegittimo' OR 'illeggittimo' OR 'illegal' OR 'non eletto') AND ('parlamento' OR 'governo' OR 'renzi' OR 'presidente')}
\end{quote}
The query is enriched with synonyms -- as in \emph{('illegittimo' OR 'illeggittimo' OR 'illegal' OR 'non eletto')} -- that take into account singular/plural forms, different jargon, and, possibly, frequent spelling errors.
With the terminology of information retrieval, these synonyms are expected to increase the recall of our filters.
On the other hand, to assess the precision of the filters we manually verified a sample of 200 tweets per filter, finding that all of them where somewhat related to the corresponding propaganda item.
The size of this sample, albeit limited, must be commensurate with the total number of tweets matching each filter, which is in the order of a few thousands.
It is worth noticing that we do not aim at perfect accuracy; rather, as any query-based filter, the goal was collecting a sufficiently large and significant sample of tweets for each propaganda item. 

In a previous work we classified these stories into four categories \cite{guarino2019beyond}, by distinguishing entirely fabricated content from manipulated items and broader propaganda pieces.
Here, we decided to focus upon the four most shared Propaganda Items (PI), and namely:
\begin{description}
\item[PI1] A newspiece about alleged vote rigging organized by government forces;
\item[PI2] A second item framing the referendum as the political product of an illegitimately elected parliament and/or government;
\item[PI3] A third news, claiming that victory of the YES would make Italy yield national sovereignty to EU institutions (especially referring to an hidden clause in art.117);
\item[PI4] A fourth - more general - piece supporting the claim that a victory of the YES would have caused a shift towards authoritarianism.
\end{description}

All the most diffused news items can be broadly located along the spectrum  of different arguments of conspiracy theories, traditionally driven by a belief that a powerful group of people is manipulating the public, while concealing their activities. As some scholars have demonstrated~\cite{castanho2017elite}, conspirationism is associated with different sub‐dimensions of populist attitudes – people‐centrism, anti‐elitism, and a good‐versus‐evil view of politics –, with coup d’état attempts and secret plots organized by political élites to gain further power or consolidate their privilege or the explicitly plot to notch the integrity of the electoral process by gaining unauthorized access to voting machines and altering voting results. 

\section{Classification of tweets and users}

After having identified the most relevant news-pieces in our dataset, we aimed at gaining a better understanding of users in our dataset and the relation between polarization and disinformation.
To classify the stance of each tweet with respect to the referendum question, we adopted a semi-automatic self-training process, described more in detail in~\cite{guarino2019beyond}.
The underlying idea is that political exchanges in social-media platforms exhibiting ``a highly partisan community structure'' with ``homogeneous clusters of users who tend to share the same political identity''~\cite {conover2011political}.
This is reflected on Twitter by the usage of different patterns of hashtags by supporters of opposite factions~\cite{becatti2019extracting}.
We therefore built a hashtag graph, selecting the top 30 hashtags by weighted degree (\emph{i.e.}, with the greatest number of co-occurrences with other hashtags).
Among them, we identified a set of generic and/or out-of-context hashtags that could have been detrimental to identifying clear and meaningful clusters, namely: ``\#referendum'', ``\#referendumcostituzionale'', ``\#photo'', ``\#riformacostituzionale'', ``\#costituzione'', ``\#4dicembre'', ``\#trendingtopic'' and ``\#1w1l''.
Pruning these hashtags indeed increased the modularity of the clustering.
The rationale was to mimic the removal of stopwords or very frequent words in order to improve the quality of topic modeling.
Louvain's algorithm was then applied to cluster such hashtags based on their mutual co-occurrence patterns.
We found the two greatest clusters to clearly identify the YES and NO fronts, thus we used hashtags in these clusters to extract a training set composed of tweets labelled as follows: $-1$ (NO) if the tweet only contains hashtags from the NO cluster; $+1$ (YES) if the tweet only contains hashtags from the YES cluster; $0$ (UNK) if the tweet contains a mix of hashtags from the two clusters.

To extend the labeling to all tweets in the dataset, we defined a text-based classifier.
The classifier may be tuned to represent tweets using tf-idf vectors, doc2vec~\cite{le2014distributed}, or a combination of both, and to use either Logistic Regression or a Gradient Boosting Classifier.
We tested any possible combination and selected the overall best performing one, namely, a Gradient Boosting Classifier using doc2vec feature vectors.
As classification score we used the mean accuracy on 10K tweets of test data and corresponding labels, with 10-fold cross-validation.
Significantly, the obtained accuracy was very high (above 90\%) and this is the reason why we did not investigate more advanced and recent classification methods.
Our explanation for this excellent accuracy is that the dichotomic nature of the referendum fostered the emergence of sets of highly partisan hashtags, rarely used in a mix.
Albeit the classifier uses the whole text of the tweets, it takes advantage of the presence of such hashtags to obtain remarkable performances.
Unfortunately, we cannot guarantee equal accuracy of our classifier on other datasets -- defining a high-quality and general purpose classifier being well beyond the scope of this paper.

On the whole, UNK tweets were substantially negligible -- although this may be due to limitations of the classifier~\cite{guarino2019beyond} -- while NO tweets were almost 1.5x more frequent than YES tweets, supporting the diffused belief that the NO front was significantly more active than its counterpart in the social debate.
Significantly, we also obtained a continuous score in [-1,1] for users, since a user can be classified with the average score of his/her tweets.
These user-level scores are used in the following sections for correlating polarization with other network properties of our corpus of users.

\begin{figure}[ht]
    \centering
    \begin{subfigure}[b]{0.47\textwidth}
        \includegraphics[width=\textwidth]{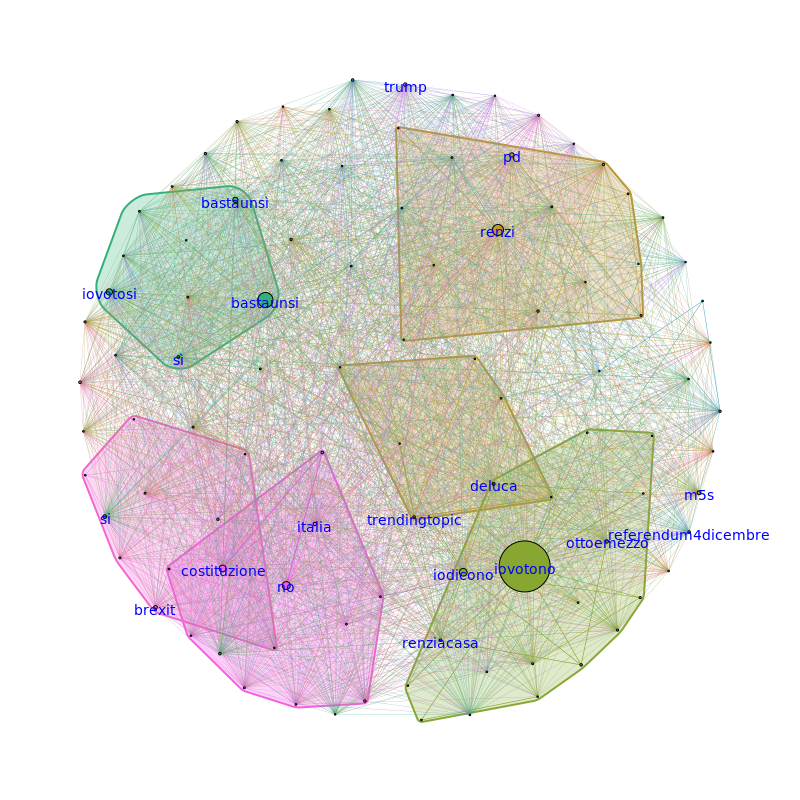}
        \caption{Hashtag graph, with clusters highlighted. Vertex size is by pagerank and top 20 hashtags are annotated.}
	    \label{fig:hashtag graph}
    \end{subfigure}
    \hfill
    \begin{subfigure}[b]{0.47\textwidth}
        \includegraphics[width=\textwidth]{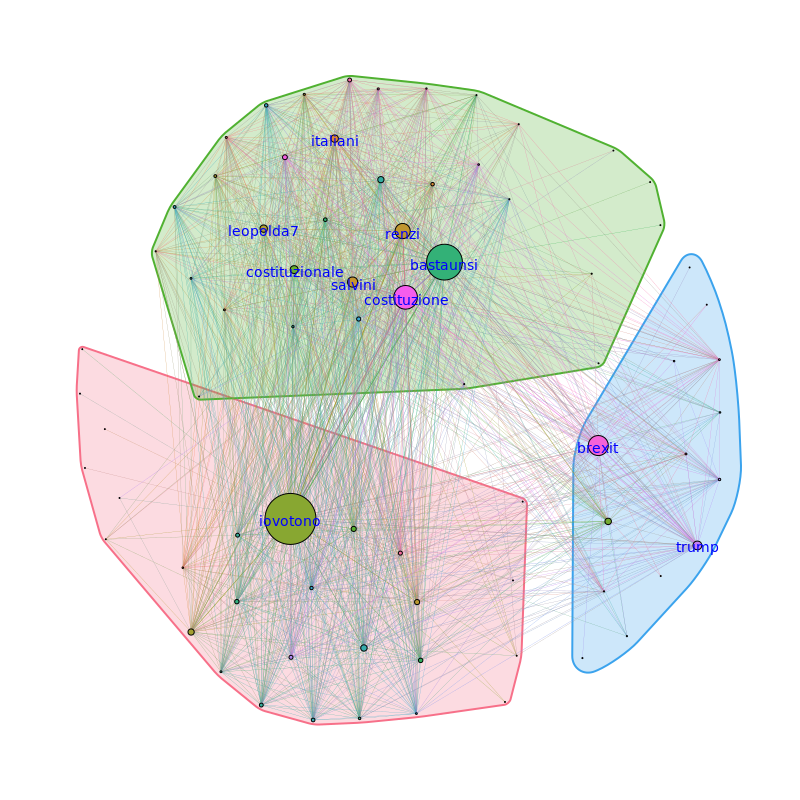}
        \caption{Cluster graph. Vertex size is by cluster size and top 10 clusters are annotated with a reference hashtag.}
	    \label{fig:clustergraph}
    \end{subfigure}
    \caption{The hashtag graph and the associated cluster graph.}
    \label{fig:clustergraphs}
\end{figure}

For the sake of clarity and completeness, the hashtag graph and its cluster-graph -- wherein each cluster is contracted into a single node -- are shown in Figure~\ref{fig:clustergraphs}.
We see that: (i) hashtags used by the NO and YES supporters are strongly clustered; (ii) ``neutral'' hashtags (such as those used by international reporters) also cluster together; (iii) a few hashtags are surprisingly high-ranked, such as ``\#ottoemezzo'', a popular political talk-show being central in the NO cluster -- thus confirming regular patterns of behavior in the ``second-screen'' use of social network sites to comment television programs ~\cite{trill2015}.
In particular, the two largest clusters of hashtags clearly characterize the two sides: the YES cluster is dominated by the hashtags “\#bastaunsì” (“a yes is enough”) and “\#iovotosi” (“I vote yes”), whereas the NO cluster by “\#iovotono” (“I vote no”), “\#iodicono” (“I say no”) and “\#renziacasa” (“Renzi go home”). 
In this perspective, the jargon of both communities show clear segregation and high levels of clustering by political alignments, as expected.

\section{Polarized retweet graphs}\label{sec:polarization}

\begin{figure}[ht]
\centering
    \includegraphics[width=.95\textwidth]{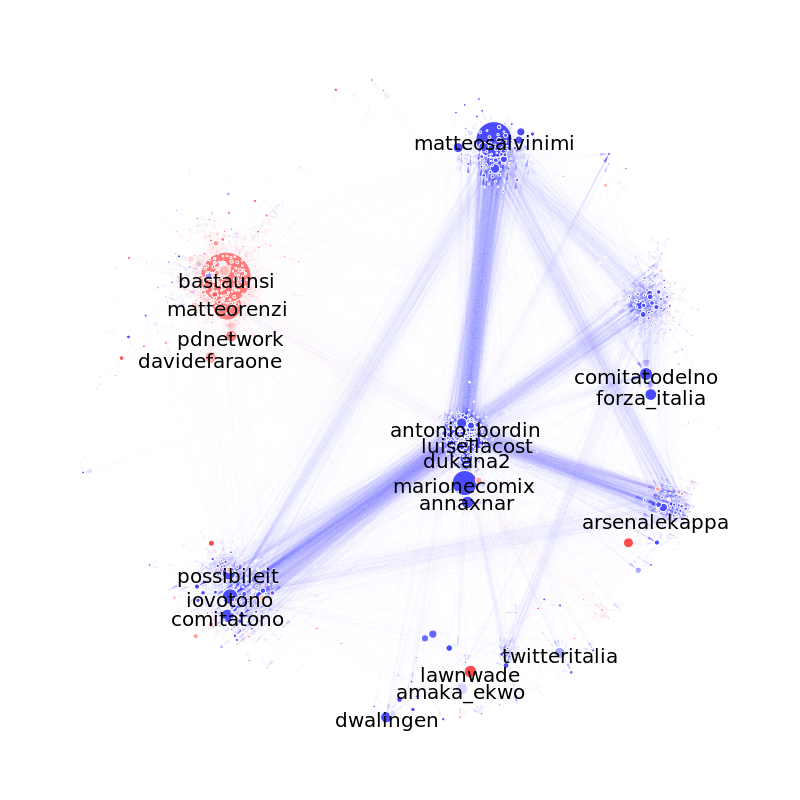}
    \caption{The top 1000 vertices of the whole graph by pagerank. Vertex size is by pagerank, vertex and edge color is by polarization (YES=red, NO=blue), the 20 top users are annotated.}
    \label{fig:graph_plot}
\end{figure}

The main objects of analysis of this paper are a set of interaction networks extracted from a Twitter dataset of more than 1.3 million tweets.
Each of these networks is formally represented as a graph $G=(V,E)$, whose vertex set $V$ models a corpus of social media users.
Specifically, as often done in the literature [49, 34], we consider directed and weighted retweet graphs, wherein nodes are Twitter users and an edge e = (u, v) means that user u retweeted user v at least once in the considered corpus of tweets.
In our graphs, edges are weighted by a parameter $w_e$ equal to the number of retweets between a given pair of users.
Nodes are instead endowed with a ``polarization'' attribute $p_u\in[-1,1]$ -- defined in the previous section -- equal to the average polarization of the tweets and retweets of that user.
Specifically, in this paper we consider the following six graphs:
\begin{itemize}
    \item The \textbf{whole} retweet graph, obtained from the entire dataset\footnote{Precisely, we only consider the giant weakly connected component of this graph, which contains 92.55\% of all vertices and 99.13\% of all edges of the complete retweet graph.}. 
    \item The \textbf{P/D} (Propaganda/Disinformation) retweet graph, obtained from the set of all tweets that matched any of the queries defined in the \nameref{sec:background}~Section, \emph{i.e.}, tweets related to any of the 12 news stories.
    \item The \textbf{PI1}, \textbf{PI2}, \textbf{PI3} and \textbf{PI4} retweet graphs, induced by the set of tweets that satisfied each of the four selected propaganda items, taken individually.
\end{itemize}
The subgraph of the whole graph composed of the 1000 vertices having greatest pagerank is shown in Figure~\ref{fig:graph_plot}.
We can clearly see a few features of the graph that will be better discussed in the following: a general prevalence of NO edges (\emph{i.e.}, tweets), multiple NO-leaning clusters and a single main YES-leaning cluster. 

A first relevant perspective on our dataset is obtained by considering how the vertex set of the P/D graph may be decomposed based on the belonging of its users to the individual PI graphs:
\begin{itemize}
    \item 67.61\% of all users in the P/D graph (\emph{i.e.}, 3666 users) are only involved in one of PI1, PI2, PI3, PI4;
    \item 16.17\% of all users in the P/D graph (\emph{i.e.}, 877 users) are involved in two of PI1, PI2, PI3, PI4;
    \item 6.79\% of all users in the P/D graph (\emph{i.e.}, 368 users) are involved in three of PI1, PI2, PI3, PI4;
    \item only 1.44\% of all users in the P/D graph (\emph{i.e.}, 78 users) are involved in all four of PI1, PI2, PI3, PI4;
    \item 7.99\% of all users in the P/D graph (\emph{i.e.}, 433 users) are involved in other PI besides PI1, PI2, PI3, PI4;
\end{itemize}
Summing up, the four items of propaganda that we selected involve approximately 92\% of all users of the P/D graph, with users only involved in other propaganda/fake news stories adding up to just 7.99\%.
We can thus safely focus on these four items without a significant loss in the generality of our results.
At the same time, the fact that most users of the P/D graph were only involved in a single PI and that only a negligible fraction was involved in all four Pis warns us of the pitfalls of considering disinformation as a whole.

\begin{figure}[ht]
\centering
    \includegraphics[width=.8\textwidth]{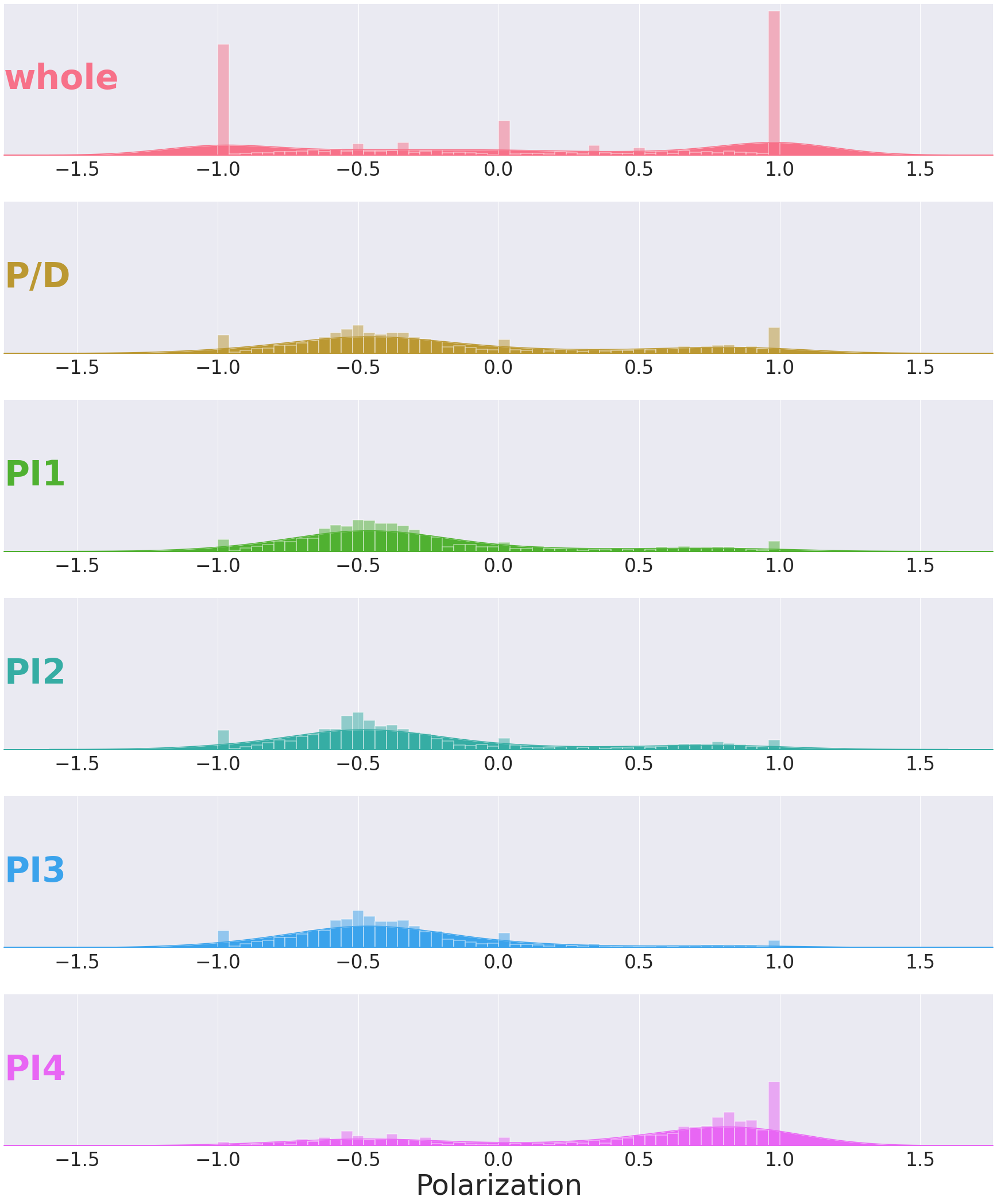}
    \caption{Polarization of users involved in propaganda.}
    \label{fig:polarization}
\end{figure}

A second aspect to consider is the distribution of the polarization attribute $p_u$ across the six graphs.
For each of the six considered graphs, Figure~\ref{fig:polarization} shows the histogram and a kernel density estimate obtained considering the value of $p_u$ for all users of the graph.
Let us remind that $p_u\in[-1, +1]$ expresses the stance of user $u$ with respect to the referendum in the range [NO,YES].
Overall, users appear to be strongly polarized, with two huge spikes at -1 and +1 for the whole graph.
When we switch to networks of propaganda, however, users seem to be generally less polarized.
This apparently counter-intuitive phenomenon is a consequence of our scoring method and of the much higher average activity of users involved in these networks.
Indeed, a user's polarization is well-definite when that user has a single tweet and gets blurrier as the number of tweets increases, because of the contribution of many tweets not all of which are necessarily equally polarized.
The average number of tweets per user in our propaganda networks is 
8 to 14 times
greater than the average computed over the whole graph.
At the same time, users with a single tweet are 37\% of the whole graph, but just 1\% to 5\% of the P/D and PIs graphs.

The distribution of PI1, PI2 and PI3 follows the overall trend of the P/D graph, that is, a general prevalence of NO users over YES users.
Since all 4 selected items, as well as most of the 12 items, are pro-NO, this may be interpreted as a prevalence of propaganda over counter-propaganda.
In that sense, PI4 is the exception: a clear example of a topic mostly used by one side (the YES coalition) to accuse the other of using deceptive propaganda.
This is a first element in favour of the importance of accounting for polarization when characterizing these propaganda networks and their users.

\section{Clustering structure}\label{sec:clustering}

\begin{figure}[htbp]
\centering
    \includegraphics[width=0.8\textwidth]{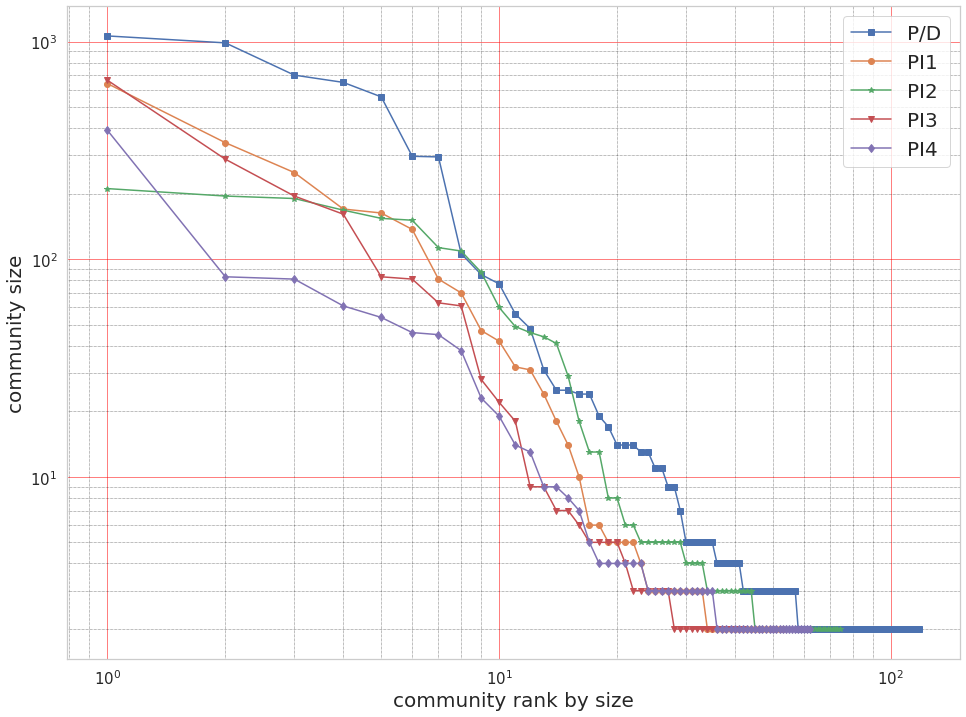}
    \caption{Comparison of the clustering structure (cluster size distribution) for different propaganda graphs.}
    \label{fig:cluster_size_distribution}
\end{figure}

The clustering structure of a retweet graph highlights relevant properties of how users and groups of users interact with each other, and of how easily information flows through the graph.
Along this line, recent work provided clear evidence that modularity-based clustering applied to retweet graphs brings to light communities of users with strong homophily/affiliation within which propaganda and polarized information spreads especially well~\cite{aragon2013communication, becatti2019extracting}.
By characterizing and comparing the clustering structure obtained for our six graphs through the well-known Louvain algorithm we expect to better understand the emergence of networks and sub-networks of propaganda and measure their persistence.
To start, in Figure~\ref{fig:cluster_size_distribution} we show the size distribution of communities for the P/D and PIs retweet graphs: 
we rank the communities of each graph based on their size and we plot the size of each community on a log scale. 
At a high level, we see that the distributions of all PIs graphs are somewhat similar -- especially for PI1 and PI3 -- and that in all cases only a few clusters have a relevant size.

\begin{figure}[htbp]
\centering
    \includegraphics[width=.95\textwidth]{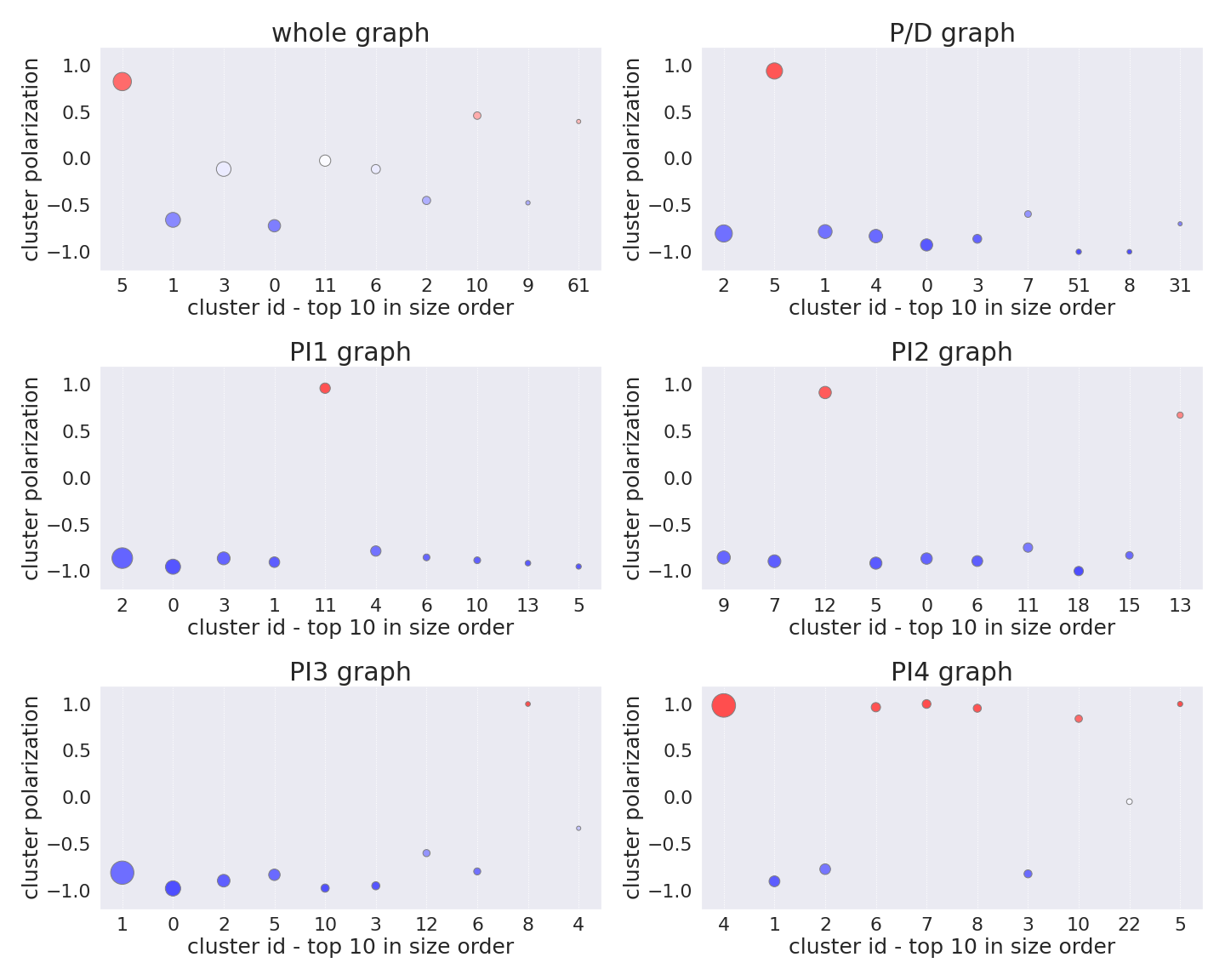}
    \caption{The polarization of the top 10 clusters for each of the six graphs. The marker size is proportional to the cluster size. The polarization is also visible from the marker color.}
    \label{fig:clusters_polarization}
\end{figure}

We now assess whether modularity based clustering detects communities of users with a clear attitude towards the referendum.
To obtain a single polarization score for a given cluster $c$, we computed the number of YES users in $c$, denoted $Y_c$, the number of NO users in $c$, $N_c$, and defined $p_c=\frac{Y_c-N_c}{Y_c+N_c}$.
This definition guarantees that $p_c=+1$ if $N_c=0$, $p_c=-1$ if $Y_c=0$ and $p_c=0$ if $Y_c=N_c$.
Yet, if compared with just taking the average polarization of the users in $c$, this measure is more robust with respect to classification accuracy -- under the assumption that telling apart YES and NO users is easier than measuring the exact polarization of each user.
In Figure~\ref{fig:clusters_polarization} we consider the 10 largest clusters of each graph ranked by size and, for each of such clusters, we plot the polarization score $p_c$.
The marker size is set proportional to the cluster size, whereas the marker color is also descriptive of the polarization in a range from blue (NO) to red (YES).
We can clearly see that the clusters of the networks of propaganda are generally and significantly more polarized than the clusters of the whole graph.
We also see that the overall prevalence of NO users in the P/D, PI1, PI2 and PI3 graphs already emerged in Figure~\ref{fig:polarization} is reflected in a greater number of NO clusters -- the same happening in PI4 for the YES front.

The main clusters of the whole graph deserve special attention.
As already observed in the literature~\cite{becatti2019extracting}, in fact, they quite clearly reflect political affiliation:
\begin{itemize}
    \item Cluster 5 ($\approx16$K members) appears to group together members and supporters of the ``Democratic Party'', including Government members (such as PM Matteo Renzi and the Minister of Reforms Maria Elena Boschi), the official YES Committee and Renzi's foundation `Leopolda', among the others.
    \item Custer 1 ($\approx11$K members) is expressive of the ``5 Star Movement'' community. Only two of the most active users (Minister Danilo Toninelli and Senator Elio Lannutti) are official party members, however, whereas the most influential actors belong to the militant base.
    \item Cluster 0 ($\approx7.5$K members) groups the members of the souverainist right, including the two politicians Matteo Salvini and Giorgia Meloni, their political parties, and a number of supporters.
    \item Cluster 2 ($\approx3.5$K members) clearly involves the ``Forza Italia'' members and advocates. 
\end{itemize}
In this context, three large and barely-polarized clusters come to light.
On the one hand, cluster 3 ($\approx10$K members) seems to validate the claim that ``structure segregation and opinion polarization share no apparent causal relationship''~\cite{prasetya2020model}.
It includes left-wing opponents to the referendum as well as several media accounts and has very low polarization (-0.04), a probable evidence of the willingness of the left-wing members of the NO alignment to maintain a cross-partisan interaction with the democrats.
On the other hand, clusters 11 and 6 ($\approx6$K and $\approx4$K members, respectively) completely escape the party affiliation logic.
Apart from @europeelects, which produces poll aggregation and election analysis in the European Union, we only found evidence of accounts belonging to international militants of the souverainist and anti-globalization movement: they are Brexit supporters, Italian pro-Trump advocates, or journalists covering such topics in their reporting activities.
    
\begin{figure}[htbp]
\centering
    \includegraphics[width=.8\textwidth]{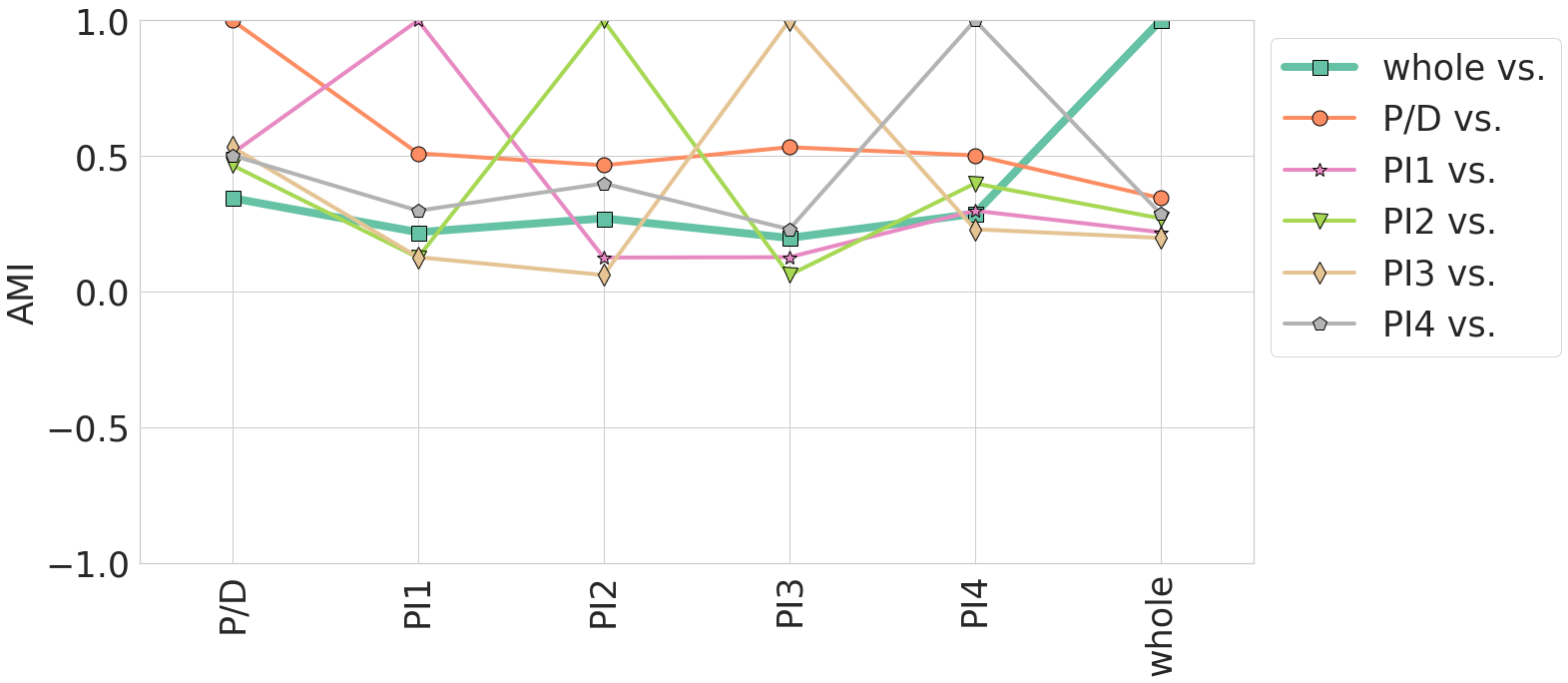}
    \caption{Pairwise Adjusted Mutual Information of graphs' clusterings.}    \label{fig:clustering_comparison}
\end{figure}

\begin{figure}[htbp]
\centering
    \includegraphics[width=\textwidth]{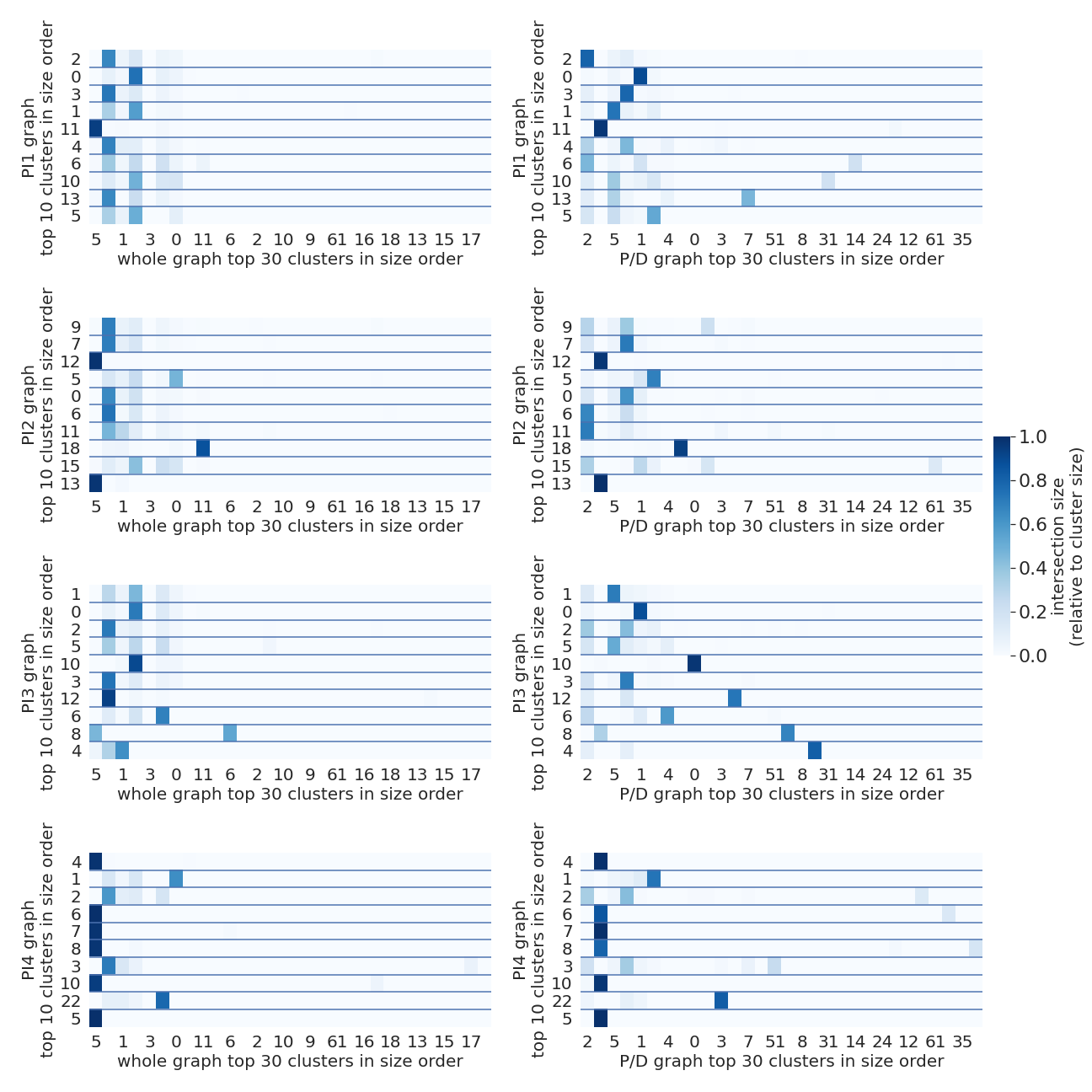}
    \caption{Cluster-to-cluster intersection size: top 10 clusters of each PI graph \emph{vs.} top 30 clusters of the whole retweet graph and the P/D graph clusters. The intersection size is normalized by the considered PI cluster size (\emph{i.e.}, each row is individually normalized and sums up to (almost) 1).}
    \label{fig:heatmap}
\end{figure}

Now, we aim at assessing to which extent the obtained clusters are influenced by the choice of a specific PI, that is, whether the patterns of cohesion among different users seem to be coherent across different topics of discussion.
In Figure~\ref{fig:clustering_comparison} we use the Adjusted Mutual Information (AMI) to compare the clusters emerged in different graphs.
Specifically, for each graph we draw a polyline showing the AMI between that graph's clustering and all other graphs' clustering.
It is worth recalling that the AMI of two partitions is 1 if the two partitions are identical, it is 0 if the mutual information of the two partitions is the expected mutual information of two random partitions\footnote{Here, the meaning of ``random'' depends on the choice of a distribution over the set of all possible partitions~\cite{vinh2009information}}, and it is negative if the mutual information of the two partitions is worse than the expected one.
Of course, when comparing the partitions obtained for any two graphs, we just consider the users that are common to both graphs.
In addition, in Figure~\ref{fig:heatmap}, we provide a more pointwise analysis of the 10 greatest communities of each PI graph, showing how users of these clusters distribute over the greatest 30 communities of the whole and P/D graphs.
Precisely, in each heatmap the cell at the intersection of row $i$ and column $j$ measures the proportion of users of cluster $i$ in the considered PI graph that lie in cluster $j$ of the compared graph.

The two figures together provide clear evidence that users are clustered in an rather unstable way, especially when we compare networks generated by individual PIs with the whole retweet graph and with each other.
The topological organization of the NO front is adequately expressive of different ideological affiliations of NO sponsors, but these differences are not clearly visible in the participation to clusters of the PI1, PI2 and PI3 networks.
Assuming that selective exposure and social validation are core driving polarization mechanisms~\cite{prasetya2020model}, two main interpretations are possible: either (i) being generally NO-leaning is enough to trigger the exposure to these three PIs, with the actual political community a user belongs to playing a marginal role; or (ii) the interactions occurring globally on Twitter -- and, as such, global information flows -- are only partially responsible of the tendency of users to diffuse propaganda and disinformation items.
In PI4, on the other hand, most clusters are \emph{de facto} sub-clusters of a macro-community of the whole retweet graph.
This is easily explained by the different polarization emerged in Figure~\ref{fig:clusters_polarization}: the macro-community is cluster 5, which we already identified as the ``YES community''.
The YES cluster seems to be driven by both an effort of community building and the attempt to de-legitimate the NO front by debunking its news-claims and propaganda items. 
As a consequence, YES users stay attached in the P/D and in other PI graphs, while they splits into sub-communities in PI4, showing a stronger degree of internal homogeneity and highlighting a polarized conversational archetype, with partisan actors and segregated community structure and discussion.

\section{Users' Centrality}
In this section we study the role played by the users in the four propaganda items selected and discussed in the previous sections. To assess the activity of each user we compute the following centrality measures on the retweets graph: PageRank, In-Degree, Out-Degree, Authority Score and Hub Score~\cite{kleinberg1999authoritative}. 
The centrality measures we chose are often used for networks analysis and their interpretation depends on the phenomenon modeled by the network. In our graph, the In-Degree tells us which are the users that are more often retweeted, \emph{i.e.}, the users creating contents that are spread on the network. The PageRank tells us which are the users that are most likely ``visited'', \emph{i.e.}, the users whose contents are most probably read if the retweets graph is used to surf the network. The Out-Degree tells us which are the users that more often retweet, \emph{i.e.}, the users playing a main role in the information diffusion.
Finally the Hub Score and Authority Score are interconnected: the former tells us which are the users that more often retweet contents created by an \emph{authority}, we call these users \emph{hubs}; the latter tells us which are the users creating the main contents about a discussion topic, we call these users \emph{authorities}.
The main difference between the Hub Score and the Out-Degree is about the content of the retweets done by a user, in the former case the user retweets authoritative contents, in the latter case the user does not show any preference about the tweet's origin. 
Similarly, the main difference between the Authority Score and the In-Degree resided in the type of users that usually retweet contents produced by a given user, in the former case the users retweeting these contents are \emph{hubs}, in the latter case there is no distinction among users.
Thus, a user has a high Authority Score if she has a high degree and the users retweeting her contents are hubs.
Tables \ref{tab:all}, \ref{tab:pi1}, \ref{tab:pi2}, \ref{tab:pi3} and \ref{tab:pi4} show for each propaganda item the top 10 users of each metric. The color of each cell denotes the polarization of the user, blue is used for NO supporters and red is used for YES supporters. The color's intensity shows how strong is the polarization, \emph{i.e.}, the darker is the color the more polarized is the user.

Our results show a few relevant aspects.
First, if we look at the whole retweet network the most active users are almost all NO supporters, as showed in Table~\ref{tab:all}, albeit the number of NO and YES users is quite balanced in the network -- as showed in Figure~\ref{fig:polarization}.
Indeed, we find only few accounts belonging to the YES supporters in the top position of the five metrics. 
Additionally, by analyzing individual propaganda items we observe that the set of most active users, their ranking and their polarization change depending on the considered PI and metrics.
In PI1, PI2 and PI3 the most active accounts are NO supporters, while in PI4 the YES supporters are the most active and numerous, in accordance with Figure~\ref{fig:polarization} and despite PI4 also being a pro-NO item. 
The analysis of users polarization is thus essential to understand the role played by the main actors inside each network: if we only considered the centrality of the users, without looking at their polarization, we would not be able to distinguish between accounts that are contributing to the diffusion of a fake news and accounts that are working against, \emph{i.e.}, the debunkers. 
We also see, again in accordance with the analysis presented in the \nameref{sec:clustering} Section, that different PIs see different users take on different roles.
Well known public figures -- such as ``mattosalvinimi'', ''giorgiameloni'' and ``matteorenzi'' -- are a minority with respect to grassroots activists, and users playing a central role in a specific network of propaganda and/or with respect to a specific metrics are absent or not as relevant in other cases -- such as ``cinmir89'' or ``proudman811''.

\begin{table}[ht]
\centering
\small 
\captionof{table}{Whole retweet graph: top 10 accounts by centrality.} 
\label{tab:all}
\begin{tabular}{lllll}
\toprule
               In-Degree            &           Out-Degree              &        Authority Score                &         Hub Score             &       PageRank\\
\midrule
\cellcolor{\NO!85} matteosalvinimi & \cellcolor{\NO!83} iovotono        & \cellcolor{\NO!22} antonio\_bordin    & \cellcolor{\NO!47} marino29b        & \cellcolor{\SI!86} bastaunsi\\
\cellcolor{\NO!22} antonio\_bordin & \cellcolor{\NO!47} marino29b       & \cellcolor{\NO!95} marionecomix       & \cellcolor{\NO!83} iovotono        & \cellcolor{\NO!85} matteosalvinimi\\
\cellcolor{\SI!86} bastaunsi       & \cellcolor{\NO!80} gincarbone      & \cellcolor{\NO!39} dukana2            & \cellcolor{\NO!38} franco\_dimuro  & \cellcolor{\SI!85} matteorenzi \\
\cellcolor{\NO!95} marionecomix    & \cellcolor{\NO!38} franco\_dimuro  & \cellcolor{\NO!83} iovotono           & \cellcolor{\NO!49} nativiitaliani  & \cellcolor{\NO!95} marionecomix\\
\cellcolor{\NO!39} dukana2         & \cellcolor{\NO!49} nativiitaliani  & \cellcolor{\NO!55} sevenseasmarina    & \cellcolor{\NO!71} luisaloffredo28 & \cellcolor{\NO!22} antonio\_bordin\\
\cellcolor{\NO!83} iovotono        & \cellcolor{\NO!36} gjscco          & \cellcolor{\NO!52} andfranchini       & \cellcolor{\NO!39} demian\_yexil   & \cellcolor{\NO!83} iovotono \\
\cellcolor{\SI!85} matteorenzi     & \cellcolor{\NO!71} luisaloffredo28 & \cellcolor{\NO!44} beatricedimadi     & \cellcolor{\NO!80} gincarbone      & \cellcolor{\NO!30} possibileit\\
\cellcolor{\NO!55} claudiodeglinn2 & \cellcolor{\NO!06} lelloesposito5  & \cellcolor{\NO!24} annaxnar           & \cellcolor{\NO!56} il\_brigante07  & \cellcolor{\NO!86} comitatono\\
\cellcolor{\NO!86} comitatono      & \cellcolor{\NO!39} demian\_yexil   & \cellcolor{\NO!45} oinot49            & \cellcolor{\NO!36} gjscco          & \cellcolor{\NO!57} comitatodelno\\
\cellcolor{\NO!55} sevenseasmarina & \cellcolor{\NO!52} giorgiomorresi  & \cellcolor{\NO!53} cremaschig         & \cellcolor{\NO!52} giorgiomorresi  & \cellcolor{\NO!39} dukana2\\
\bottomrule
\end{tabular}
\end{table}

%%
%% PI1 - Vote rigging - FRAUD0 - ST1
%%
\begin{table}
\centering
\small
\captionof{table}{PI1 retweet graph: top 10 accounts by centrality.}
\label{tab:pi1}
\begin{tabular}{lllll}
\toprule
           In-Degree &           Out-Degree &        Authority Score &         Hub Score  &      PageRank\\
\midrule
 \cellcolor{\NO!22} antonio\_bordin     & \cellcolor{\NO!49} nativiitaliani     & \cellcolor{\NO!22} antonio\_bordin    & \cellcolor{\NO!49} nativiitaliani  & \cellcolor{\NO!22} antonio\_bordin\\
 \cellcolor{\NO!85} matteosalvinimi     & \cellcolor{\NO!47} marino29b          & \cellcolor{\NO!85} matteosalvinimi    & \cellcolor{\NO!52} giorgiomorresi  & \cellcolor{\SI!89} didimiero\\
 \cellcolor{\NO!39} dukana2             & \cellcolor{\NO!56} il\_brigante07     & \cellcolor{\NO!39} dukana2            & \cellcolor{\NO!59} cinmir89        & \cellcolor{\NO!85} matteosalvinimi\\
 \cellcolor{\NO!25} francotrax          & \cellcolor{\NO!42} celestinoceles7    & \cellcolor{\NO!25} francotrax         & \cellcolor{\NO!53} andreazanettin  & \cellcolor{\NO!25} francotrax\\
 \cellcolor{\NO!69} carloalterego       & \cellcolor{\NO!48} proudman811        & \cellcolor{\NO!69} carloalterego      & \cellcolor{\NO!56} il\_brigante07  & \cellcolor{\NO!39} dukana2\\
 \cellcolor{\NO!55} claudiodeglinn2     & \cellcolor{\NO!6} lelloesposito5     & \cellcolor{\NO!18} eliolannutti        & \cellcolor{\NO!41} cretellaroberta & \cellcolor{\NO!18} eliolannutti\\
 \cellcolor{\NO!18} eliolannutti        & \cellcolor{\NO!41} marobe997          & \cellcolor{\NO!53} newsinunclick      & \cellcolor{\NO!10} soloio0509      & \cellcolor{\NO!57} penelopy2000\\
 \cellcolor{\NO!80} 5bc32772e3fb467     & \cellcolor{\NO!38} franco\_dimuro     & \cellcolor{\NO!44} possidonio\_gg     & \cellcolor{\NO!59} archidevivaio   & \cellcolor{\NO!23} adrimcmlxi\\
 \cellcolor{\NO!40} patriziarametta     & \cellcolor{\NO!52} giorgiomorresi     & \cellcolor{\NO!80} 5bc32772e3fb467    & \cellcolor{\NO!35} dopiot          & \cellcolor{\NO!68} ipredicatore\\
 \cellcolor{\NO!68} ipredicatore        & \cellcolor{\NO!52} kirumakataossi1    & \cellcolor{\NO!55} claudiodeglinn2    & \cellcolor{\NO!47} marino29b       & \cellcolor{\NO!60} toscaross\\
\bottomrule
\end{tabular}
\end{table}

%%
%% PI2 - Illegitimate government - PROPG0 - ST2
%%
\begin{table}
\centering
\small
\captionof{table}{PI2 retweet graph: top 10 accounts by centrality.} 
\label{tab:pi2}
\begin{tabular}{lllll}
\toprule
           In-Degree &           Out-Degree &        Authority Score &         Hub Score    &       PageRank\\
\midrule
 \cellcolor{\NO!39} dukana2         & \cellcolor{\NO!83} iovotono       & \cellcolor{\NO!39} dukana2            & \cellcolor{\NO!47} marino29b       & \cellcolor{\NO!57} comitatodelno\\
 \cellcolor{\SI!85} pdnetwork       & \cellcolor{\NO!47} marino29b      & \cellcolor{\NO!55} sevenseasmarina    & \cellcolor{\NO!49} nativiitaliani  & \cellcolor{\NO!33} renatobrunetta\\
 \cellcolor{\NO!57} comitatodelno   & \cellcolor{\NO!49} nativiitaliani & \cellcolor{\NO!22} antonio\_bordin    & \cellcolor{\NO!48} proudman811     & \cellcolor{\NO!55} sevenseasmarina\\
 \cellcolor{\NO!55} sevenseasmarina & \cellcolor{\NO!56} il\_brigante07 & \cellcolor{\NO!56} ermannokilgore     & \cellcolor{\NO!83} iovotono        & \cellcolor{\NO!39} dukana2\\
 \cellcolor{\NO!22} antonio\_bordin & \cellcolor{\NO!52} uleprr         & \cellcolor{\NO!43} fmcastaldo         & \cellcolor{\NO!41} cretellaroberta & \cellcolor{\SI!85} pdnetwork\\
 \cellcolor{\NO!92} advalita        & \cellcolor{\NO!80} gincarbone     & \cellcolor{\NO!57} comitatodelno      & \cellcolor{\NO!36} gjscco          & \cellcolor{\NO!22} antonio\_bordin\\
 \cellcolor{\NO!56} ermannokilgore  & \cellcolor{\NO!42} battistabd     & \cellcolor{\NO!41} rossellafidanza    & \cellcolor{\NO!52} uleprr          & \cellcolor{\NO!92} advalita\\
 \cellcolor{\NO!43} fmcastaldo      & \cellcolor{\NO!38} franco\_dimuro & \cellcolor{\NO!26} deboramau          & \cellcolor{\NO!56} il\_brigante07  & \cellcolor{\NO!56} ermannokilgore\\
 \cellcolor{\NO!41} rossellafidanza & \cellcolor{\NO!38} mad13021966    & \cellcolor{\NO!92} advalita           & \cellcolor{\NO!42} battistabd      & \cellcolor{\NO!42} inarratore\\
 \cellcolor{\NO!26} deboramau       & \cellcolor{\NO!48} proudman811    & \cellcolor{\NO!24} annaxnar           & \cellcolor{\NO!53} cocchi2a        & \cellcolor{\NO!43} fmcastaldo\\
\bottomrule
\end{tabular}
\end{table}

%%
%% PI3 - Italy cedes sovereignty to EU - CONSQ1 - ST3
%%
\begin{table}
\centering
\small
\captionof{table}{PI3 retweet graph: top 10 accounts by centrality.} 
\label{tab:pi3}
\begin{tabular}{lllll}
\toprule
           In-Degree &           Out-Degree &        Authority Score &         Hub Score &       PageRank\\
\midrule
 \cellcolor{\NO!55} claudiodeglinn2 & \cellcolor{\NO!47} marino29b          & \cellcolor{\NO!55} claudiodeglinn2    & \cellcolor{\NO!49} nativiitaliani  & \cellcolor{\NO!55} claudiodeglinn2\\
 \cellcolor{\NO!85} matteosalvinimi & \cellcolor{\NO!80} gincarbone         & \cellcolor{\NO!85} matteosalvinimi    & \cellcolor{\NO!52} giorgiomorresi  & \cellcolor{\NO!70} angelosica1965\\
 \cellcolor{\NO!71} luisaloffredo28 & \cellcolor{\NO!49} nativiitaliani     & \cellcolor{\NO!67} patriotail         & \cellcolor{\NO!43} mania48mania53  & \cellcolor{\NO!85} matteosalvinimi\\
 \cellcolor{\NO!67} patriotail      & \cellcolor{\NO!71} luisaloffredo28    & \cellcolor{\NO!71} luisaloffredo28    & \cellcolor{\NO!80} gincarbone      & \cellcolor{\NO!55} sevenseasmarina\\
 \cellcolor{\NO!55} sevenseasmarina & \cellcolor{\NO!52} giorgiomorresi     & \cellcolor{\NO!65} deglclaudio        & \cellcolor{\NO!68} caspanistefania & \cellcolor{\NO!65} deglclaudio\\
 \cellcolor{\NO!37} giorgiameloni   & \cellcolor{\NO!46} malaspinadavide    & \cellcolor{\NO!10} civico21           & \cellcolor{\NO!45} pietrof70       & \cellcolor{\NO!67} patriotail\\
 \cellcolor{\NO!39} liberatilinda   & \cellcolor{\NO!23} piras\_zia         & \cellcolor{\NO!70} angelosica1965     & \cellcolor{\NO!62} ilpellicano88   & \cellcolor{\NO!42} xmeridio78\\
 \cellcolor{\NO!10} civico21        & \cellcolor{\NO!47} ori254             & \cellcolor{\NO!55} sevenseasmarina    & \cellcolor{\NO!42} celestinoceles7 & \cellcolor{\NO!10} civico21\\
 \cellcolor{\NO!65} deglclaudio     & \cellcolor{\NO!42} celestinoceles7    & \cellcolor{\NO!40} carmentpf          & \cellcolor{\NO!59} archidevivaio   & \cellcolor{\NO!39} liberatilinda\\
 \cellcolor{\NO!80} 5bc32772e3fb467 & \cellcolor{\NO!55} claudiodeglinn2    & \cellcolor{\NO!40} valy\_s            & \cellcolor{\NO!57} gidal\_randagio & \cellcolor{\NO!60} toscaross\\
\bottomrule
\end{tabular}
\end{table}

%%
%% PI4 - Risk of authoritarian drift - CONSQ2 - ST4
%%
\begin{table}
\centering
\small
\captionof{table}{PI4 retweet graph: top 10 accounts by centrality.}
\label{tab:pi4}
\begin{tabular}{lllll}
\toprule
           In-Degree &           Out-Degree &        Authority Score &         Hub Score  &       PageRank\\
\midrule
 \cellcolor{\SI!86} bastaunsi       & \cellcolor{\SI!78} danieledvpd        & \cellcolor{\SI!86} bastaunsi          & \cellcolor{\SI!78} danieledvpd     & \cellcolor{\SI!86} bastaunsi\\
 \cellcolor{\SI!80} fnicodemo       & \cellcolor{\SI!79} angelinascanu      & \cellcolor{\SI!80} fnicodemo          & \cellcolor{\SI!81} rtgovernorenzi  & \cellcolor{\NO!33} renatobrunetta\\
 \cellcolor{\NO!33} renatobrunetta  & \cellcolor{\SI!78} amtomarchio        & \cellcolor{\SI!85} thelambkin\_       & \cellcolor{\SI!77} giordanobattini & \cellcolor{\NO!64} ilmattinale\\
 \cellcolor{\SI!85} thelambkin\_    & \cellcolor{\SI!81} rtgovernorenzi     & \cellcolor{\NO!28} magdazanonii       & \cellcolor{\SI!81} alcinx          & \cellcolor{\SI!80} fnicodemo\\
 \cellcolor{\NO!28} magdazanonii    & \cellcolor{\SI!83} italiarecord       & \cellcolor{\SI!62} piercamillo        & \cellcolor{\SI!86} albertoforesti3 & \cellcolor{\NO!60} fi\_online\_\\
 \cellcolor{\NO!53} paolocristallo  & \cellcolor{\SI!79} lcungi             & \cellcolor{\SI!52} belpassijessica    & \cellcolor{\SI!83} italiarecord    & \cellcolor{\NO!24} renatapolverini\\
 \cellcolor{\SI!54} eugeniocardi    & \cellcolor{\SI!84} mursino71          & \cellcolor{\SI!81} serracchiani       & \cellcolor{\SI!78} amtomarchio     & \cellcolor{\SI!85} thelambkin\_\\
 \cellcolor{\SI!62} piercamillo     & \cellcolor{\SI!81} alcinx             & \cellcolor{\SI!54} eugeniocardi       & \cellcolor{\SI!84} alfuturosi      & \cellcolor{\NO!28} magdazanonii\\
 \cellcolor{\SI!40} arsenalekappa   & \cellcolor{\SI!86} albertoforesti3    & \cellcolor{\SI!72} diegozardini       & \cellcolor{\SI!79} angelinascanu   & \cellcolor{\NO!57} comitatodelno\\
 \cellcolor{\NO!64} ilmattinale     & \cellcolor{\SI!77} giordanobattini    & \cellcolor{\SI!88} unitaonline        & \cellcolor{\SI!64} ruiciccio       & \cellcolor{\NO!53} paolocristallo\\
\bottomrule
\end{tabular}
\end{table}

\begin{figure}[htbp]
    \centering
    \begin{subfigure}[b]{0.315\textwidth}
	    \label{fig:indegSpearman}
        \includegraphics[width=\textwidth,trim={70pt 0 20pt 0},clip]{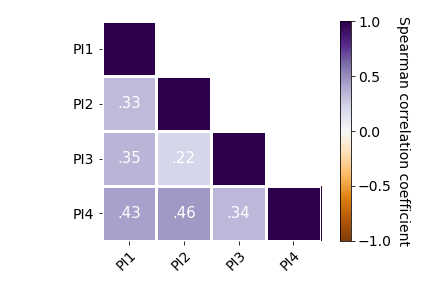}
         \caption{In-Degree [ALL]}
    \end{subfigure}
    \begin{subfigure}[b]{0.315\textwidth}
        \includegraphics[width=\textwidth,trim={70pt 0 20pt 0},clip]{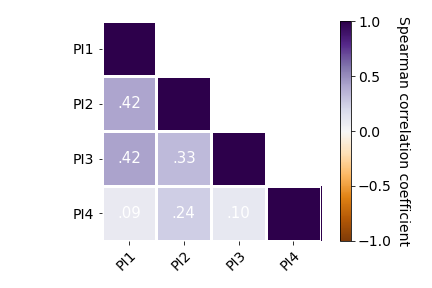}
        \caption{Out-Degree [ALL]}
        \label{fig:outdegSpearman}
    \end{subfigure}
        \begin{subfigure}[b]{0.315\textwidth}
        \includegraphics[width=\textwidth,trim={70pt 0 20pt 0},clip]{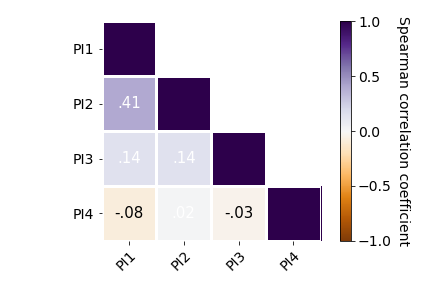}
        \caption{Authority Score [ALL]}
	    \label{fig:authscoreSpearman}
    \end{subfigure}
        \begin{subfigure}[b]{0.315\textwidth}
        \includegraphics[width=\textwidth,trim={70pt 0 20pt 0},clip]{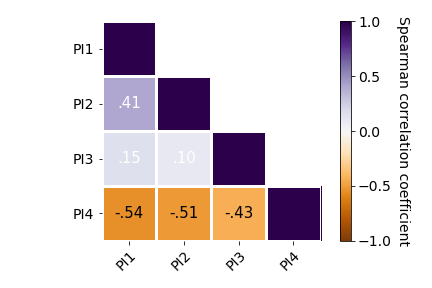}
        \caption{Hub Score [ALL]}
	    \label{fig:hubscoreSpearman}
    \end{subfigure}
        \begin{subfigure}[b]{0.315\textwidth}
        \includegraphics[width=\textwidth,trim={70pt 0 20pt 0},clip]{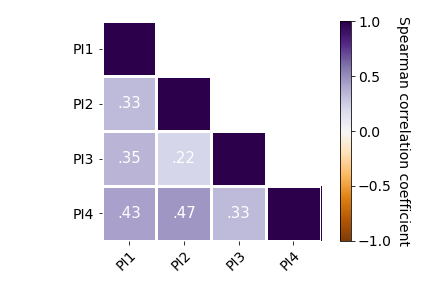}
        \caption{Pagerank [ALL]}
	    \label{fig:pagerankSpearman}
    \end{subfigure}
    \begin{subfigure}[b]{0.315\textwidth}
        \includegraphics[width=\textwidth,trim={70pt 0 20pt 0},clip]{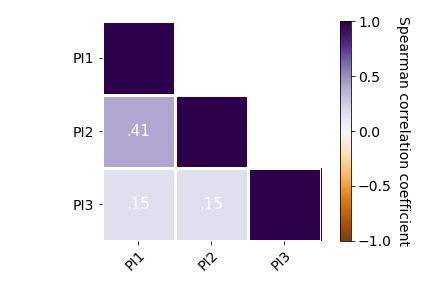}
        \caption{Authority Score [NO]}
	    \label{fig:authscoreNoSpearman}
    \end{subfigure}
    \begin{subfigure}[b]{0.315\textwidth}
        \includegraphics[width=\textwidth,trim={70pt 0 20pt 0},clip]{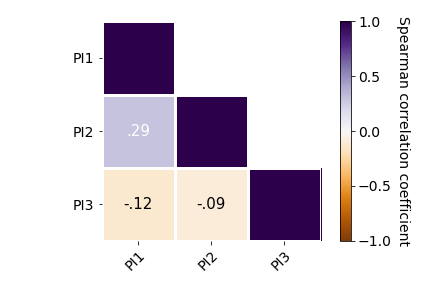}
        \caption{Authority Score [YES]}
	    \label{fig:authscoreYesSpearman}
    \end{subfigure}
    \begin{subfigure}[b]{0.315\textwidth}
        \includegraphics[width=\textwidth,trim={70pt 0 20pt 0},clip]{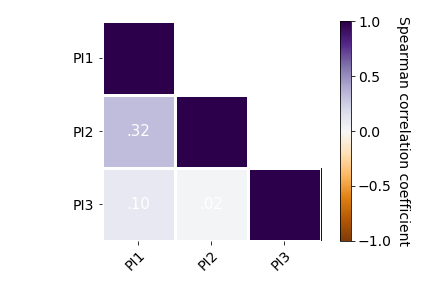}
        \caption{Hub Score [NO]}
	    \label{fig:hubscoreNoSpearman}
    \end{subfigure}   
    \begin{subfigure}[b]{0.315\textwidth}
        \includegraphics[width=\textwidth,trim={70pt 0 20pt 0},clip]{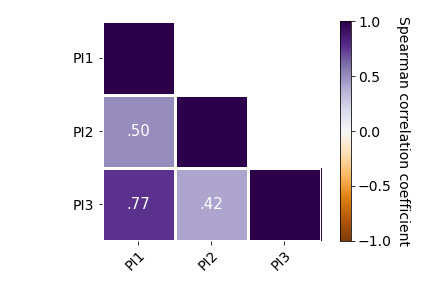}
        \caption{Hub Score [YES]}
	    \label{fig:hubscoreYesSpearman}
    \end{subfigure}    
    
    \caption{Spearman's rank correlation coefficients.}
    \label{fig:correlationSpearman}
\end{figure}

To further investigate the persistence of these rankings across different networks of propaganda, in Figure~\ref{fig:correlationSpearman} we present a set of correlation matrices that broadly corroborate the previous findings.
Specifically, for each centrality measure we report the pairwise correlation between the rankings produced by that measure on different graphs, in order to better understand the role of the users that were active in more than one propaganda item.
We rely on Spearman's rank correlation coefficient, rather than the widely used Pearson's, because we are neither especially interested in verifying linear dependence, nor we do expect to find it.
We are more interested in the possible monotonic relationship between centrality measures as determined by Spearman's correlation.

As already observed, combining the centrality and polarization data, we notice that in the PI4 network there is a different community of users that is active and that is spreading information with respect to the other PI networks.
This behaviour is clearly visible from the Hub Score matrix (Figure~\ref{fig:hubscoreSpearman}) and partially from the Out-Degree matrix (Figure~\ref{fig:outdegSpearman}). 
In the former the anti-correlation in row PI4 shows the existence of a different community of spreaders in PI4 with respect to other PI. A community composed of users that are absent, less active or play a different role in other networks.
In the Out-Degree matrix we have almost no correlation in row PI4 except for PI2. This difference is due to the presence of a small, not negligible, community of YES supporters in PI2 as showed in Figure~\ref{fig:clusters_polarization}.
Whereas, the PageRank and In-Degree matrices show that the relevance of the accounts creating contents is more stable than those diffusing the information.
Finally, the Authority Score matrix shows that, although the contents creator accounts are stable, their role change in the network.
The same account is considered more authoritative in one network and less in the other.

What is happening in the other networks can be better understood by looking at figures~\ref{fig:authscoreNoSpearman}-\ref{fig:hubscoreYesSpearman} where we computed separately the correlation for NO and YES supporters for the Authority and Hub Scores, that overall appear to be the most informative. 
To better focus our analysis we excluded the PI4 row.
Our results show that among YES supporters the content creators accounts are not stable and their role change depending on the propaganda item selected.
On the other hand, the role of the accounts spreading information is more stable, meaning that for different networks there are different authorities, but the hubs are the same. 
For what concerns the NO supporters we have that both authorities and hubs relevance changes depending on the propaganda network.
Thus there is probably a more efficient synergy among NO supporters between authority and hub accounts.

\section{Conclusions}

The paper aimed at providing new insights into the dynamics of propaganda networks on Twitter. 
The results of our study are partly in line with existing research.  Modularity-based clustering, applied to retweet graphs, pictured a wide panorama of communities of users with strong homophily/affiliation and polarized position.
As expected, the clusters of propaganda networks were generally and significantly more polarized than the clusters of the whole graph and the topological organization proved to be highly representative of the ideological affiliation of users. 
The comparison between clusters in different graphs reveals that users' clusters are rather dynamic, particularly when comparing networks generated by individual propaganda items with the whole retweet graph and with each other. 
It seems that global clusters, often associated with information exposure, are only partially responsible of the tendency of users to diffuse propaganda and disinformation items.
When it comes to taking a position on a controversial topic, users tend to group with different people with respect to those they usually connect to in the whole graph, and the ``high-level'' polarization of a user -- such as the NO \emph{vs.} YES leaning in our case -- may have a more prominent role than his/her political affiliation.
This is especially visible for users involved in propaganda – as opposed to counter-propaganda.

The combined analysis of cluster-to-cluster intersections and centrality metrics additionally indicates how different propaganda items are associated to different users with authoritative roles.
The correlation of centrality metrics across different networks provides further insights: (i) the Authority and Hub Score seem the most informative metrics for studying networks of propaganda, thanks to their ability to tell apart content creators and spreaders; (ii) the role of content creators is taken by different users for different propaganda items, independently of clusters polarization; (iii) spreaders are instead generally more ``consistent''. 
Overall, the propaganda community depicted in this study, far from being monolithic, has a considerable degree of internal variability, in terms of central actors, topics and opinion polarization.
Polarization with respect to a main theme, transversal to the considered propaganda items, emerged as a fundamental parameter in governing users behavior.

A side result of the present paper is the identification of a few expedients and precautions to be used in practice.
For instance, we showed that the authority and hub scores unveil different players of a propaganda network, and that real-time detection of propaganda and disinformation campaigns must be built on top of a reliable polarization measure.
To this end, it must be kept in mind that users' polarization (on a specific issue) and political partisanship do not always coincide: we showed that the topic of debate may significantly alter the community structure of an interaction network, and thus the perceived affiliation of its users.
Further directions of research could involve other clustering algorithms as well as dynamic influence metrics, in order to gain deeper knowledge on the relationship between exposure to propaganda and the general structure of users interaction. 

Another issue we explicitly chose not to cover involves the determinants of user centrality in a debate (why or how a user gained a central role?), nor to detect coordinated bot attacks that possibly boosted the centrality of a Twitter profile.
We rather focus on the \emph{perceived} centrality of a user, regardless of what caused it, to show that: (i) the centrality itself is of limited use if not accompanied with a polarization analysis, \emph{e.g.}, to distinguish propaganda from counter-propaganda/debunking; (ii) using different metrics make it possible to detect different roles in the network, and such roles vary from one disinformation item to another.
That said, the analysis highlighted evidence of a major coordination effort in the NO front, which is where the considered propaganda and disinformation items were more prevalent.
Understanding whether this coordination was supported by bots is left to future work.

\section*{Abbreviations}
AMI: Adjusted Mutual Information;
P/D: Propaganda/Disinformation;
PI: Propaganda Item;
PM: Prime Minister;
UNK: Unknown.

\subsubsection*{Authors' contributions}
SG, NT, ACe and ACh designed the study. ACh acquired the data. SG and ACe created the software used to perform the data analysis. SG, NT and ACe interpreted the results and wrote the paper. All authors revised the work and read and approved the final manuscript.

\subsection*{Availability of data and materials}
Part of the code used in this paper will be included in the network analysis toolbox DisInfoNet, currently under development by the partners of the Project ``SOMA'' at https://gitlab.com/s.guarino/disinfonet.
DisInfoNet is presented in a previous conference paper~\cite{guarino2019beyond} and will be released by the end of the SOMA Project.
The entire dataset used during the current study is not publicly available due to Twitter's policies.
The ids of the tweets are available from the corresponding author on reasonable request.

\subsection*{Funding}
This work was supported in part by the Project ``SOMA'', funded by the European Union's Horizon 2020 research and innovation programme under grant agreement No 825469.
The European Commission had no role in the design of the study and collection, analysis, and interpretation of data and in writing the manuscript.
Any opinion, finding, and conclusions expressed in this paper only reflect the views of the authors.

\bibliographystyle{unsrt}
\bibliography{biblio}

\end{document}